\documentclass[twocolumn]{aastex61}

\usepackage{natbib}
\usepackage{graphicx}
\usepackage{amssymb}
\usepackage{multirow}
\usepackage{float}

\defcitealias{HeinkeHo2010}{HH10} 	
\defcitealias{Ho2009}{HH09} 	
\defcitealias{Elshamouty2013}{E+13}
\defcitealias{Posselt2013}{P+13}

\newcommand{\be}{\begin{equation}}
\newcommand{\ee}{\end{equation}}

\shorttitle{Upper limits on the rapid cooling of the Cas A CCO}
\shortauthors{Posselt \& Pavlov}

\begin{document}
 
\title{Upper limits on the rapid cooling of the Central Compact Object in Cas A}
\author{B. Posselt}
\correspondingauthor{B.Posselt}
\affil{Department of Astronomy \& Astrophysics, Pennsylvania State University, 525 Davey Lab,University Park, PA 16802, USA}
\email{posselt@psu.edu}

\author{G. G. Pavlov}
\affil{Department of Astronomy \& Astrophysics, Pennsylvania State University, 525 Davey Lab,University Park, PA 16802, USA}

\begin{abstract}
The Central Compact Object (CCO) in the Cassiopeia A supernova remnant is most likely a very young ($\approx 300$\,yr) neutron star.
If a previously reported decrease of its surface temperature by 4\% in 10 years could be confirmed, it would have profound theoretical implications for neutron star physics. However, the temperature decrease was inferred from \emph{Chandra} ACIS data affected by instrumental effects which could cause time-dependent  spectral distortions. Employing a different instrument setup which minimizes 
spectral distortions, our 2006 and 2012 \emph{Chandra} spectra of the CCO did not show a statistically significant temperature decrease.
Here, we present additional observations from 2015 taken in the same instrument mode.
During  the time span of 8.5 years, we detect no significant temperature decrease, using either carbon or hydrogen atmosphere models in the X-ray spectral fits.
Our conservative $3\sigma$ upper limits
correspond to $<3.3$\% and $<2.4$\%
temperature decrease
in 10 years 
for carbon atmosphere model fits with varying or constant
values of the absorbing hydrogen column density, respectively.
The recently revised model for the ACIS filter contaminant has a strong effect on the fit results, reducing the significance of
the previously reported temperature and flux changes. 
We expect that a further improved contaminant model and longer time coverage can significantly lower the upper limits in the future.
\end{abstract}


\section{Introduction}
\label{intro}
One of the methods to investigate the composition, structure and physical properties in the interior of neutron stars is to study their thermal evolution (e.g., \citealt{Page2004, Yakovlev2004}).
Using \emph{Chandra} observations of the Central Compact Object (CCO) in the Cassiopeia A (Cas A) supernova remnant, and fitting the CCO spectrum with a carbon atmosphere model, \citet[HH10 hereafter]{HeinkeHo2010} reported an unexpectedly rapid 4\% ($5.4\sigma$) decline of the surface temperature and a 21\% flux decline over the time span of 10 years. 
This rapid cooling was interpreted by \citet{Shternin2011} and \citet{Page2011} as due to enhanced neutrino emission caused by the recent onset of neutron superfluidity (formation of Cooper pairs) in the neutron star core. 
Considered as the first direct evidence that superfluidity and superconductivity occur in superdense matter of neutron stars, this result has been widely discussed (over 100  publications in  2011-2018).
However, the rapid cooling was inferred from \emph{Chandra} ACIS-S Graded mode observations that suffered from several instrumental effects. 
The most important one is photon pileup, where two or more photons are detected as a single event\footnote{For more details, see \url{cxc.harvard.edu/ciao/ahelp/acis_pileup.html}}. Pileup can distort the observed CCO spectrum. The pileup fraction in the observed spectrum of a given constant source decreases over time because the decreasing sensitivity of the ACIS detector.
This is mostly due to an accumulating contaminant on the optical-blocking filters of the ACIS detectors. In addition, not all X-ray events are telemetered in the Graded mode\footnote{For more details, see \url{cxc.harvard.edu/ciao/why/cti.html}}, potentially also affecting the spectrum.
For these reasons, the spectral changes reported by HH10 required confirmation.
Analyzing observations with different \emph{Chandra} instruments, 
\citet[E+13 in the following]{Elshamouty2013} reported a statistically significant decrease again only in the case of the ACIS-S  Graded mode observations where the best-fit decay was $3.5\%\pm 0.4$\% (from 2000 to 2010). 
Avoiding the spectral distortion effects caused by photon pileup and the use of the Graded telemetry mode, \citet{PavlovL2009} and \citet{Posselt2013} (P+13 in the following) employed a more suitable instrument mode of ACIS-S in 2006 and 2012 to probe the spectral evolution of the Cas A CCO.
Using hydrogen and carbon atmosphere models, P+13 reported that the statistical significance of any temperature change between 2006 and 2012 did not exceed $2.5\sigma$, at the default calibration, for all the considered constraints on the fitting parameters. However, the time coverage was only six years, and the uncertainties were too large to completely rule out the previously reported ``rapid cooling''. Here, we report on new observations with the same ACIS instrument mode, extending the time coverage of the CCO monitoring with this more suitable ACIS observing mode to $8.5$ years.\\ 

The accumulating ACIS contaminant complicates the analysis of all the CCO \emph{Chandra} ACIS data.
Errors in the contamination correction can lead to an offset of the derived spectral parameters from the correct values.
P+13 evaluated the influence of the time-variable optical depth of the contaminant on the spectral fit results. 
An imperfect contamination correction impacts the inferred absorbing hydrogen column density, $N_{\rm H}$,  which in turn is correlated with the inferred temperature. 
Based on the apparently increasing best-fit values of $N_{\rm H}$ from 2006 to 2012, P+13 speculated that an underestimated optical depth of the ACIS contamination could explain any remaining spectral changes of the Cas A CCO.
The ACIS Calibration team developed a new model for the ACIS contamination \citep{Plucinsky2016} which is available since December 2016.\footnote{\url{cxc.harvard.edu/caldb/downloads/Release_notes/CALDB\_v4.7.3.html\#TD\_ACIS\_CONTAM\_10}}
Here, we also report on the effect of this new contamination model on the results from the 2006 and 2012 subarray data.\\    

It is important to emphasize that all the temperature changes were only found if a carbon atmosphere model was used to describe the surface emission of the CCO. Only for this atmosphere model the assumption of a constant emission area is reasonable. This assumption, however, imposes an additional constraint on the spectral fit. When P+13 allowed the emission area to vary, the result was an (insignificantly) \emph{increased} temperature for both carbon and hydrogen atmosphere models. As P+13 showed, hydrogen atmosphere models fit the CCO spectra equally well. The currently existing X-ray data do not allow one to differentiate between these two atmosphere models (see also \citealt{Alford2017}). The carbon atmosphere model was preferred by HH10 and E+13 because it implies an emission size consistent with what one would expect for the entire surface of a neutron star, while the hydrogen atmosphere models results in a much smaller emission area. This could indicate the presence of one or more hot spots. However, one would expect X-ray pulsations in such a case, but none have been found so far (see, e.g., \citealt{PavlovL2009,Mereghetti2002}) though the derived upper limits on the pulsed fraction are above the values measured for other CCOs (e.g., \citealt{Gotthelf2013} and references therein).

\begin{figure*}[]
{\includegraphics[width=170mm]{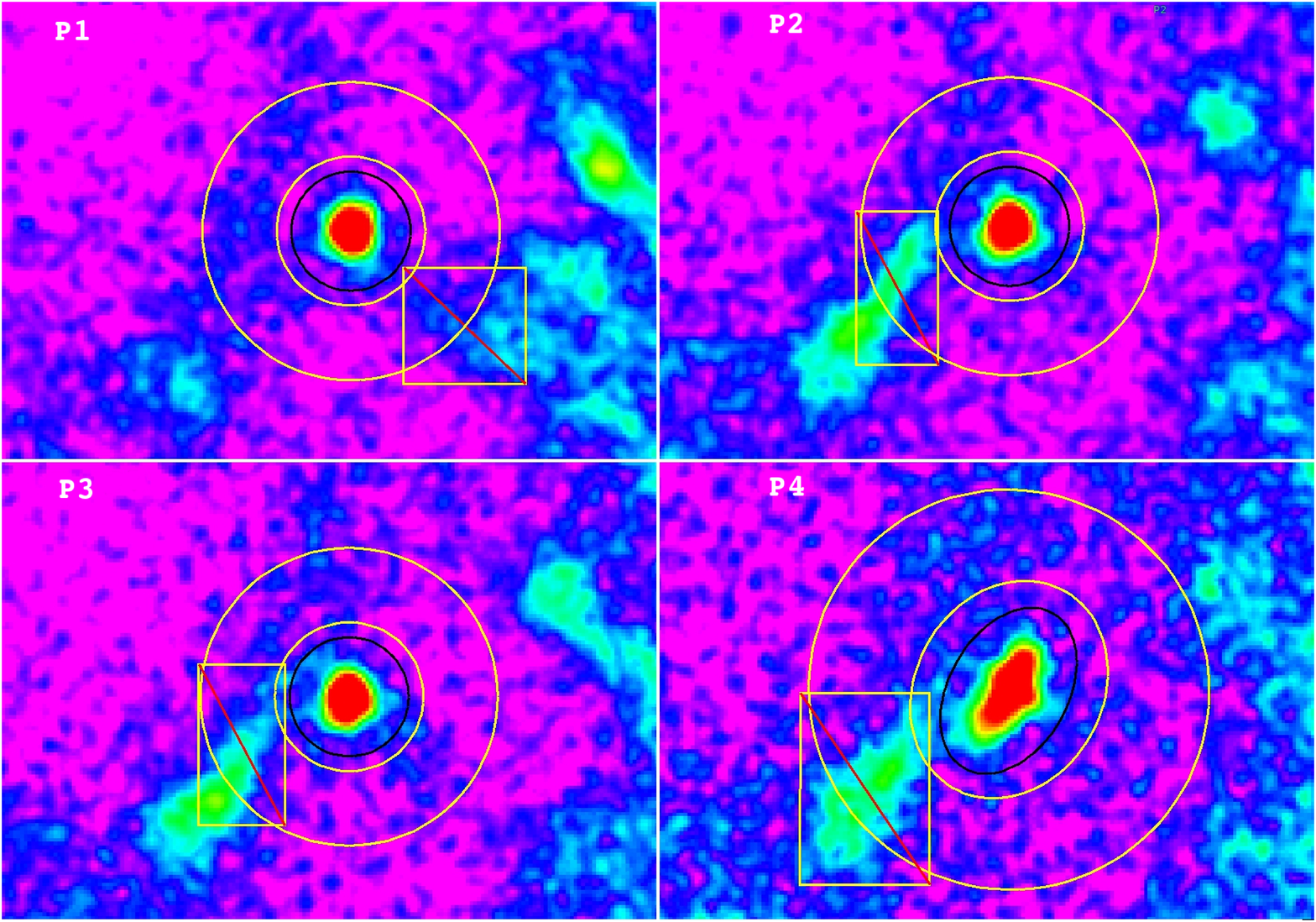}}\\ 
\caption{Images of the CCO and its vicinity from observing epochs P1 (2006), P2 (2012), P3 (2015), and P4 (2015, three days after P3). North is up, East is to the left, all the images at the same spatial scale. Marked in each image are: the inner source extraction region in black (radius of 4 ACIS pixels $1\farcs{97}$ in P1-P3), the annulus region for the background in yellow lines (inner radius of 5 pixels, outer radius of 10 pixels in P1-P3), and the box regions excluded from the respective background regions in yellow. In P4, the CCO looks elongated because it is located at an off-axis angle of $190\arcsec$ from the optical axis.}
\label{ExtractRegions}
\end{figure*}

\begin{figure}[t]
{\includegraphics[height=85mm, angle=90]{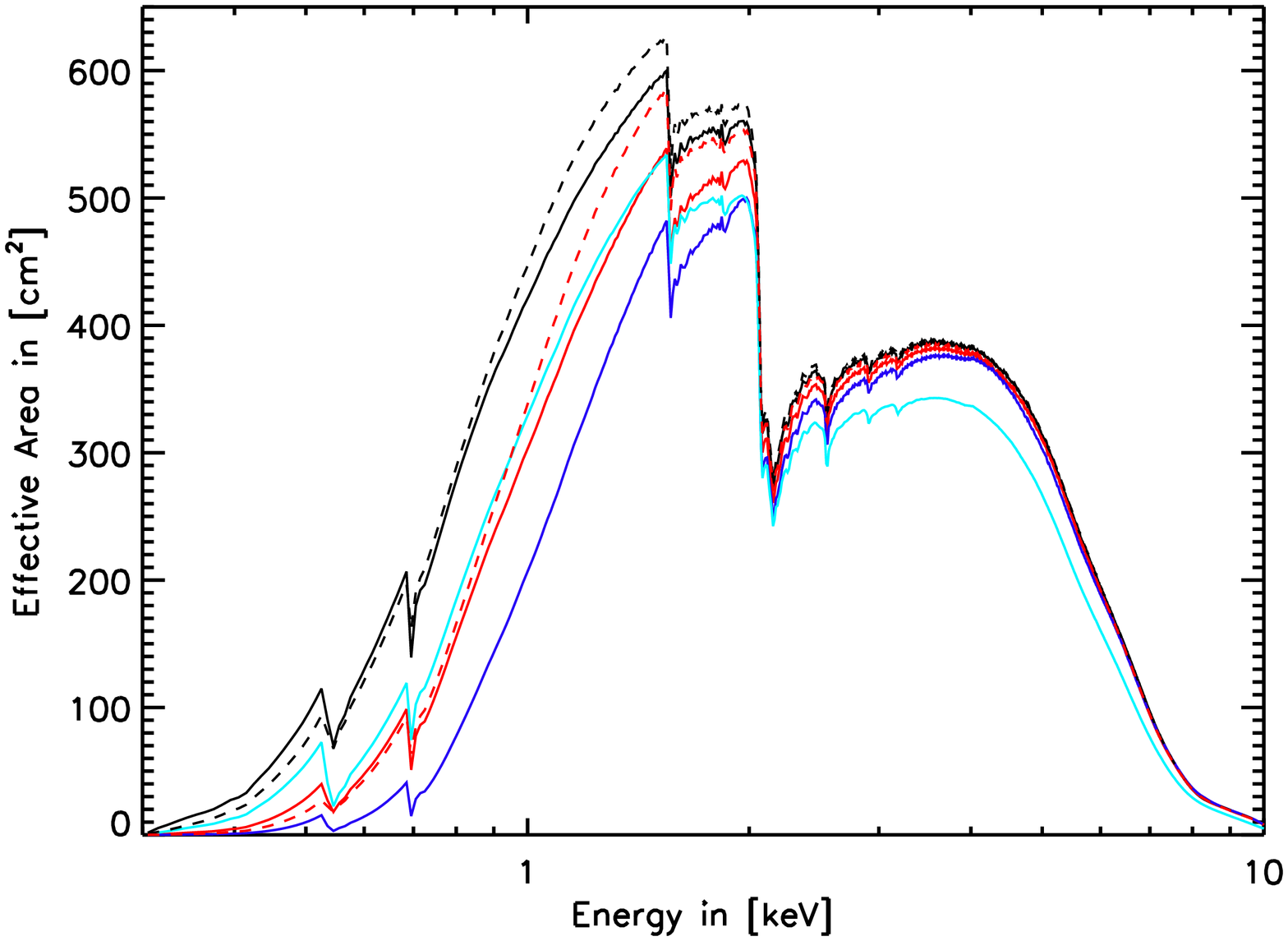}} 
\caption{Effective areas at the position of the CCO target on the ACIS-S3 chip for different epochs:
P1 -- black, P2 -- red, P3 -- blue, P4 (center of the chip) -- cyan. The solid lines correspond to the calibration CALDB 4.7.3, the dashed lines to CALDB 4.5.5.1 which was available for our previous work (P+13). It is clearly seen that the contamination for the old data was underestimated compared to the new contamination model.}
\label{effis}
\end{figure}

\section{Observations and Data Reduction}
\label{obsred}
In order to derive the most stringent constraints on the potential spectral evolution of the Cas A CCO, we use only \emph{Chandra} ACIS subarray mode observations. In subarray mode, only a part of the ACIS chip is read out, allowing a quicker registration of the X-ray events and a substantial reduction of the pileup for bright sources. 
In each observation (listed in Table~\ref{table:obs}) the target was imaged on the ACIS-S3 chip in the 100 pixel subarray.
This reduces the frame time to 0.34\,s versus the 3.24\,s in full-frame mode, reducing the pile-up fraction to less than 1.6\% in all epochs (in comparison to $\sim 20$\,\% in the case of the full frame mode, \citealt{PavlovL2009}).
The analysis of observing epochs P1 (2006) and P2 from (2012) was presented by P+13, but these data are re-analyzed here using updated calibration data. 
Observing epochs P3 and P4 are three days apart in May 2015. Similar to P1 and P2, the subarray was placed near the chip readout in P3 to reduce the charge transfer ineffiency effect on the spectrum.
For P4,  the subarray was placed at the center of the ACIS chip in order to probe for effects due to a nonuniform distribution of the contaminant over the chip. This observation was taken to facilitate a re-calibration of the Graded mode observation (the analysis and results of which will be presented elsewhere). The comparison of P3 and P4 provided a valuable check of the consistency of the ACIS calibration since the CCO is unlikely to change its emission properties within three days.\\ 

The measured offsets between the CCO centroid positions in P1--P4 are insignificant if the absolute astrometry uncertainty of \emph{Chandra}, $0\farcs{4}$, is taken into account. 
We extracted source and background regions as indicated in Figure~\ref{ExtractRegions}.
Intervening filaments of the supernova remnant are excluded from the background regions. We note that the extraction regions for P1 and P2 are the same as used by P+13. Spectra are binned with a signal to noise ratio of at least 10.\\

We employed several different versions of CIAO and CALDB during the course of our analysis because we noticed that updates on the ACIS filter contamination  correction were changing the results.
The comparison of the analysis results from P3 and P4, for example, initially showed significant differences in the derived spectral fit parameters even though these observations were taken just three days apart in the same instrument mode, albeit at different chip positions. 
A major update on the contamination model was implemented in CALDB version 4.7.3.
The new contamination model version 10 resulted in a substantial improvement in the consistency of the results obtained for P3 and P4.
Later updates on the CALDB (tested up to version 4.7.7) did not result in any noticable changes in the effective areas or fit results for P1 to P4.
In the following, all results are given for CIAO version 4.9 \citep{Fruscione2006} with CALDB version 4.7.3.
The spectral analysis was carried out with XSPEC (version 12.8.2, \citealt{Arnaud1996}).\\

\begin{deluxetable}{lrllllrrr}[]
\vspace{0.1cm}
\tablecaption{Observation Parameters \label{table:obs}}
\tablewidth{0pt}
\tablehead{
\colhead{ID} & \colhead{ObsID} & \colhead{MJD} & \colhead{$T_{\rm exp}$} & \colhead{$C$} & \colhead{$f_{\rm Src}$} & \colhead{S3$_X$} & \colhead{S3$_Y$} &  \colhead{$\theta$}\\
\colhead{} & \colhead{} & \colhead{days} & \colhead{ks} & \colhead{cts} & \colhead{\%} & \colhead{pix} & \colhead{pix} & \colhead{$\arcsec$}\\
}
\startdata
P1  &  6690 & 54027 & 61.6 & 7441 & 86.4 & 211.2 & 49.4 & 7.8\\
P2  & 13783 & 56052 & 63.4 & 6720 & 87.4 & 215.5 & 51.5 & 7.3\\
P3  & 16946 & 57140 & 68.1 & 6278 & 87.8 & 229.6 & 55.2 & 23.0\\
P4  & 17639 & 57143 & 42.7 & 4562 & 82.4 & 574.84 & 509.03 & 190.4\\
\enddata
\tablecomments{The ID indicates the abbreviation used for the observing epoch of the \emph{Chandra} data set with the listed ObsID, $T_{\rm exp}$ is the dead-time-corrected exposure time after GTI filtering, (total) counts $C$ in the energy range $0.3-6$\,keV and the source count fraction $f_{\rm Src}$ correspond to the source extraction regions in Figure~\ref{ExtractRegions} and spectral fits in Table~\ref{table:carbonfits}. S3$_X$ and S3$_Y$ are the centroid chip coordinates on ACIS-S3. $\theta$ is the off-axis angle.} 
\end{deluxetable}

The goal of this paper is to give an update on the possible spectral and thermal evolution of the Cas A CCO. 
Since the spectral differences obtained from previous data sets had an impact on the inferred thermal evolution only if the spectrum was modeled with a carbon atmosphere model, we concentrate our analysis on this model. We emphasize, however, that a hydrogen atmosphere fits the data equally well (as we verified using the complete subarray data set), but requires an emission area smaller than the total neutron star surface (see P+13 for a detailed discussion). 
We also restrict the presented results to the case of fixed normalization,
$\mathcal{N}=R^2_{\rm NS} / d_{\rm 10kpc}^2$, where  $R_{\rm NS}$ is the assumed neutron star radius in km, and $d_{\rm 10kpc}$ is the distance in 10\,kpc, because  
the significance of the temperature (or flux) \emph{difference} was shown to be very similar to those obtained using tied, but free normalizations (assuming the same emission size in different epochs) (P+13).
As in P+13, we use the carbon atmosphere models by \citet{Suleimanov2014} with a surface gravitational acceleration of $\log g=14.45$ and a gravitational redshift of $z=0.375$, which corresponds to a neutron star with $M_{\rm NS}=1.647$\,M$_{\odot}$ and  $R_{\rm NS}=10.33$\,km.
We use a distance of 3.4\,kpc ($d=3.4^{+0.3}_{-0.1}$\,kpc; \citealt{Reed1995}).

\begin{deluxetable*}{lllllc}[]
\tablecaption{Fit results for the carbon atmosphere models with $\log g=14.45$ and $z=0.375$\protect\\  \label{table:goddt}}
\tablewidth{0pt}
\tablehead{
\colhead{Data} &  \colhead{$N_{\rm H}$} & \colhead{$T_{\rm eff}$} & \colhead{$F^{\rm{abs}}_{-13}$} & \colhead{$F^{\rm{unabs}}_{-12}$}  & \colhead{$\chi^2_{\nu}$/dof} \\
\colhead{ }  & \colhead{$10^{22}$\,cm$^{-2}$} &  \colhead{$10^4$ K} & \colhead{ } & \colhead{ } & \colhead{}   
}
\startdata
\tableline
P3  &  $2.15 \pm 0.08$ & $198.3^{+1.4}_{-1.3}$ &  $6.96^{+0.18}_{-0.17} $  & $2.69 \pm 0.12 $ & 1.065/84\\[0.7ex]
P4  &  $2.10 \pm 0.09$ & $198.5 \pm 1.8$ &  $7.11 \pm 0.23$  &  $2.71^{+0.16}_{-0.15}$ & 1.065/84 \\[0.7ex]
\tableline
P3  &  $2.13 \pm 0.06$ & $198.0 \pm 1.2$ &  $6.95 \pm 0.17$  & $2.67 \pm 0.10 $ & 1.056/85\\[0.7ex]
P4  &  $=N_{\rm H}$ in P3 & $198.9 \pm 1.4$ &  $7.14 \pm 0.22$ & $2.74 \pm 0.12$  & 1.056/85 \\[0.7ex]
\tableline
\tableline
\enddata
\tablecomments{The fits were done simultaneously for P3 and P4. The normalization is fixed at $\mathcal{N}=R^2_{\rm NS} / d_{\rm 10kpc}^2=923$ (see text), where  $R_{\rm NS}$ is the assumed neutron star radius in km, and $d_{\rm 10kpc}$ is the distance in 10\,kpc.
Fluxes are given for the energy range of 0.6-6\,keV. $F^{\rm{abs}}_{-13}$ is the absorbed flux in units of $10^{-13}$\,erg\,cm$^{-2}$\,s$^{-1}$, while $F^{\rm{unabs}}_{-12}$ is the unabsorbed flux in units of $10^{-12}$\,erg\,cm$^{-2}$\,s$^{-1}$. The reduced $\chi^2_{\nu}$ and the degrees of freedom (dof) of the fit are listed in the last column.
All errors indicate the 90\% confidence level for one parameter of interest.}
\end{deluxetable*}

\section{Results and Discussion}

\begin{figure}[]
{\includegraphics[width=80mm, bb=25 12 570 540, clip]{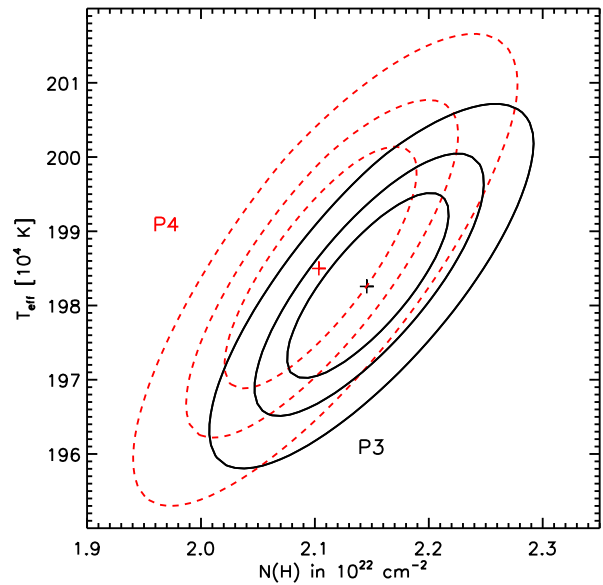}} 
\caption{Temperature versus $N_{\rm H}$ confidence contours (68\,\%, 90\,\%, 99\,\%) for the fit to the carbon atmosphere model with $\log g=14.45$, $z=0.375$. Note that in this and all the following contour plots we mark the contour levels for two parameters of interest.
In the spectral model for this and the following figures, it is assumed that the whole surface of a neutron star is emitting in X-rays and that the distance is 3.4\,kpc (norms fixed). The black contours mark P3 where the CCO position is located 55 pixel away from the chip boundary, the red contours mark P4 (3 days after P3) where the CCO is located in the center of the $1024\times 1024$\,pix chip.
\label{carbonNHTgoddt}}
\end{figure}

First, we compare the results from the two observations in 2015, P3 and P4. As shown in Table~\ref{table:goddt} and Figure~\ref{carbonNHTgoddt}, the fit values for the two epochs are consistent with each other within their 90\% confidence levels.
We note that if CALDB versions earlier than version 4.7.3 were used, then the fit values were significantly offset, in particular those of $N_{\rm H}$. 
Although the differences are insignificant, we note that the higher $N_{\rm H}$ value is found in P3, where the ACIS filter contamination at the target position (close to CCD readout) is expected to be higher than in P4\footnote{See Figure 3 in "The Spatial structure in the ACIS OBF contamination memo" 2004,  \url{http://hea-www.harvard.edu/~alexey/acis/memos/cont_spat.pdf}}. 
Because the temperature and $N_{\rm H}$ are correlated fit parameters, forcing $N_{\rm H}$ to be the same in P3 and P4 increases the temperature difference. Since the best-fit values still agree within their $1\sigma$ uncertainties, we utilize the increased count statistics in 2015 and tie the parameters of P3 and P4 in all following fits. We fit all epochs simultaneously and obtain the results listed in Table~\ref{table:carbonfits}.\\

\begin{figure}[]
{\includegraphics[width=80mm, bb=25 12 570 540, clip]{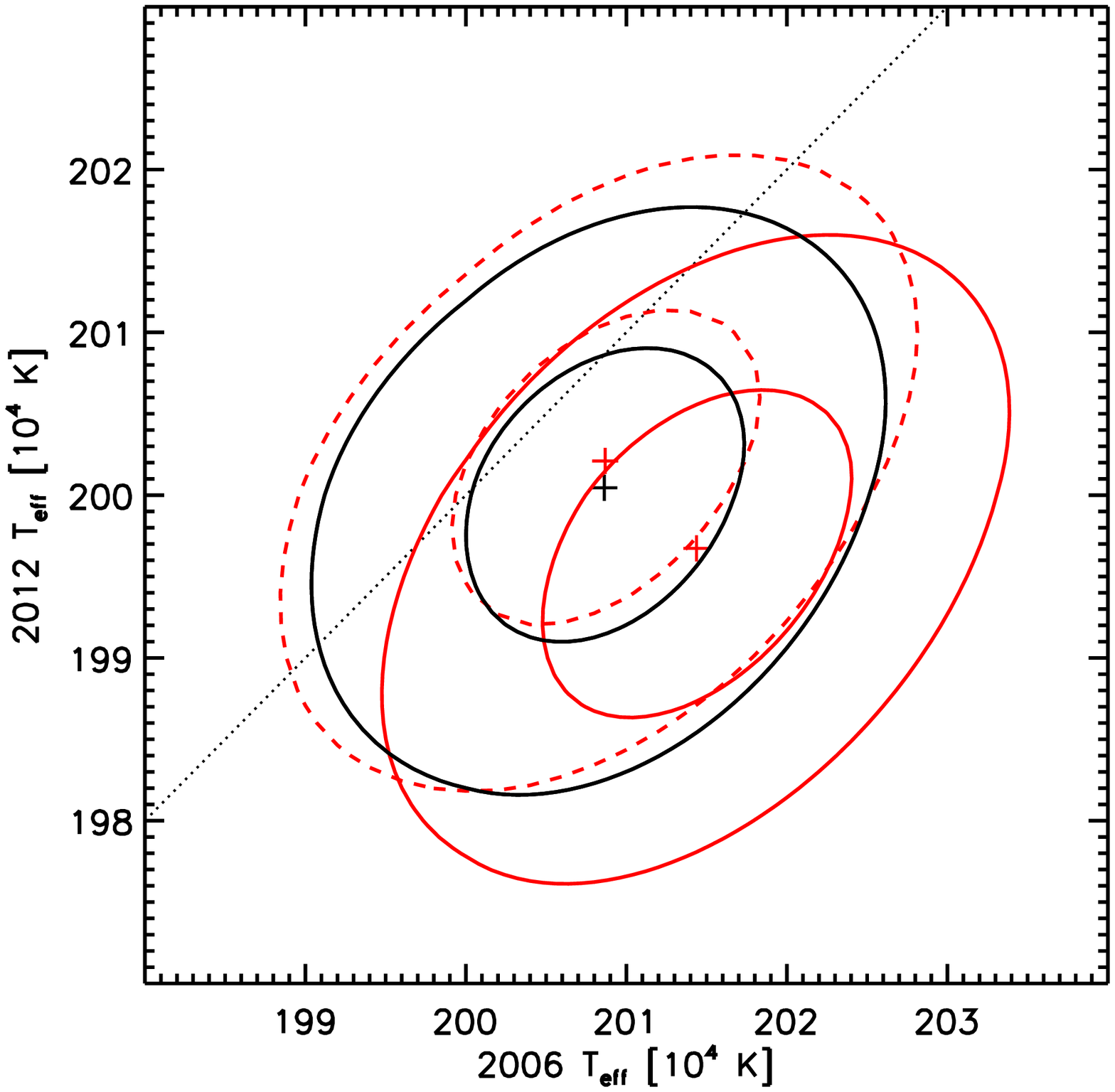}}
\caption{The temperature confidence contours (68\,\%,  99\,\%) of P1 and P2 for the updated calibration (ACIS filter contamination model 10, CALDB 4.7.3) are shown with black solid contours.  For this fit, $N_{\rm H}$ is set to be the same in all epochs (see Table~\ref{table:carbonfits} for the fit results). 
The red solid contours mark the result obtained with CALDB 4.5.5.1, the red dashed contours correspond to the shifted (old) contours if the contamination layer is crudely approximated to be \emph{under}estimated by 30\% (all red contours from P+13). The temperature difference between 2006 and 2012 is less significant with the new contamination model.
\label{t1t2w2013}\vspace{-0.2cm}}
\end{figure}

Next, we compare our new results with those of P+13. 
The temperature (and flux) difference between P1 and P2 has become even less significant due to the new contamination model, see
Table~\ref{table:carbonfits} and Figure~\ref{t1t2w2013}. This is true for the fit with $N_{\rm H}$ set to be the same in all epochs as well as if it is allowed to vary between the different observing years (though not between P3 and P4). This result illustrates that a spatially and temporally accurate ACIS contamination model is indeed crucial for the study of the potential temperature change of the Cas A CCO.\\

\begin{figure}[b]
\centering
{\includegraphics[height=85mm, angle=270, clip]{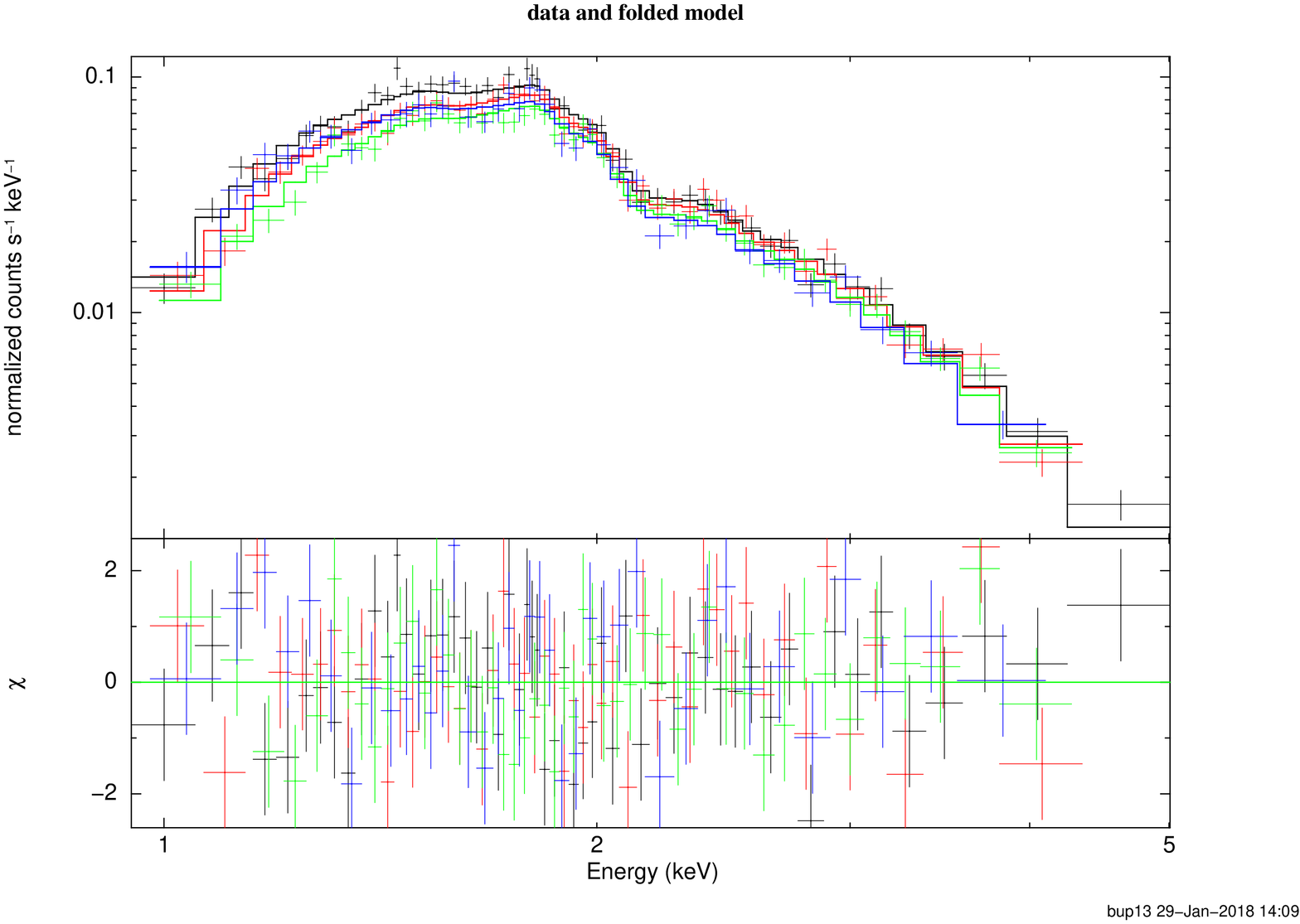}} 
\caption{The data and our fit to the carbon atmosphere model with $\log g=14.45$, $z=0.375$, see Table~\ref{table:carbonfits} for the fit results. $N_{\rm H}$ values are allowed vary between epochs. The lower panel shows the fit residuals in units of sigmas. Black, red, green, blue correspond to epochs P1, P2, P3, and P4, respectively.
\label{carbondatres}}
\end{figure}

\begin{deluxetable*}{llllllc}[t]
\tablecaption{Fit results for the carbon atmosphere models with $\log g=14.45$ and $z=0.375$ \label{table:carbonfits}}
\tablewidth{0pt}
\tablehead{
\colhead{Data}  & \colhead{$N_{\rm H}$} & \colhead{$T_{\rm eff}$} &  \colhead{$F^{\rm{abs}}_{-13}$} & \colhead{$F^{\rm{unabs}}_{-12}$}  & \colhead{$L_{\rm bol}^{\infty}$} & \colhead{$\chi^2_{\nu}$/dof} \\
\colhead{ }  & \colhead{$10^{22}$\,cm$^{-2}$} &  \colhead{$10^4$ K} & \colhead{ } & \colhead{ } & \colhead{$10^{33}$\,erg\,s$^{-1}$} & \colhead{}   
}
\startdata
P1      & $2.16 \pm 0.04$ & $200.9\pm 0.9$ & $7.49 \pm 0.17$  & $2.88 \pm 0.09$ & $6.5 \pm 0.1$ & 1.06/190\\ [0.7ex]
P2      & $=N_H$(P1) & $200.5^{+0.9}_{-1.0}$ & $7.30 \pm 0.17$  & $2.82 \pm 0.08 $ & $6.4 \pm 0.1$ & 1.06/190 \\[0.7ex]
P3\&P4  & $=N_H$(P1) & $198.8 \pm 0.9$  & $7.05 \pm 0.14$  & $2.74 \pm 0.07 $ & $6.3 \pm 0.1$ & 1.06/190 \\[0.7ex]
$\triangle$ (P1 -- P2), 5.54\,yr    & & $-0.8 \pm 1.1$ & $-0.19^{+0.23}_{-0.22} $  & $-0.06 \pm 0.09$ \\
$\triangle$ (P2 -- P3\&P4), 2.98\,yr & & $-1.2^{+1.5}_{-1.4} $ & $-0.24 \pm 0.20$  & $-0.08 \pm 0.08$ \\
$\triangle$ (P1 -- P3\&P4), 8.53\,yr & & $-2.0 \pm 1.0$ & $-0.44^{+0.21}_{-0.20}$  & $-0.14 \pm 0.08$ \\
\hline
P1  &  $2.18 \pm 0.06$ & $201.2 \pm {1.2}$ & $7.52 \pm 0.20$  & $2.91 \pm 0.08 $ &  $6.6 \pm 0.2$ & 1.06/188\\[0.7ex]
P2  &  $2.19 \pm 0.07$ & $200.4^{+1.2}_{-1.3} $ & $7.34 \pm 0.20$ & $2.84\pm 0.08$ &  $6.5 \pm 0.2$ & 1.06/188 \\[0.7ex]
P3\&P4  & $2.13\pm 0.06$ & $198.3 \pm 1.1$  & $7.06 \pm 0.15$  & $2.68^{+0.07}_{-0.06} $ & $6.2 \pm 0.1$ & 1.06/188 \\[0.7ex]
$\triangle$ (P1 -- P2), 5.54\,yr &    & $-0.8 \pm 1.7$ & $-0.18^{+0.28}_{-0.27} $  & $-0.07 \pm 0.12$ \\
$\triangle$ (P2 -- P3\&P4), 2.98\,yr & & $-2.1^{+1.8}_{-1.7}$ & $-0.27 \pm 0.24$  & $-0.16^{+0.11}_{-0.10} $ \\
$\triangle$ (P1 -- P3\&P4), 8.53\,yr & & $-2.9 \pm 1.6$ & $-0.45 \pm 0.25$  & $-0.23 \pm 0.10$ \\
\enddata
\tablecomments{The fits were done simultanously for P1-P4, the parameters are tied for P3 and P4. The normalization is fixed for all epochs in all fits at $\mathcal{N}=923$ (see text).
Fluxes are given for the energy range of 0.6-6\,keV. $F^{\rm{abs}}_{-13}$ is the absorbed flux in units of $10^{-13}$\,erg\,cm$^{-2}$\,s$^{-1}$, while $F^{\rm{unabs}}_{-12}$ is the unabsorbed flux in units of $10^{-12}$\,erg\,cm$^{-2}$\,s$^{-1}$. 
All errors indicate the 90\% confidence level for one parameter of interest. The uncertainties of the differences are obtained from the contour plots of the respective two parameters of interest, e.g., Figure~\ref{t2t1t2t3}.
The bolometric luminosity at inifinity is calculated as $L^{\infty}_{\rm bol}=4 \pi \sigma {R^{\infty}_{\rm Em}}^2 {T^{\infty}_{\rm eff}}^4= 4 \pi \sigma 10^{10} \mathcal{N} d_{\rm 10kpc}^2 {T_{\rm eff}}^4 (1+z)^{-2}$\,erg\,s$^{-1}$. Its uncertainty only considers the uncertainty of the temperature.\vspace{-0.5cm}}
\end{deluxetable*}

As we noted above, the fit results in Table~\ref{table:carbonfits} were derived by tying the fit parameters for the data of P3 and P4, since these observations were taken just three days apart. The exposure-weighted average observing date is MJD 57141.2, which we will use as the time of epoch 3 in the following. 
The data and the fit for a variable $N_{\rm H}$ are shown in Figure~\ref{carbondatres}.\\

\begin{figure}
{\includegraphics[width=85mm]{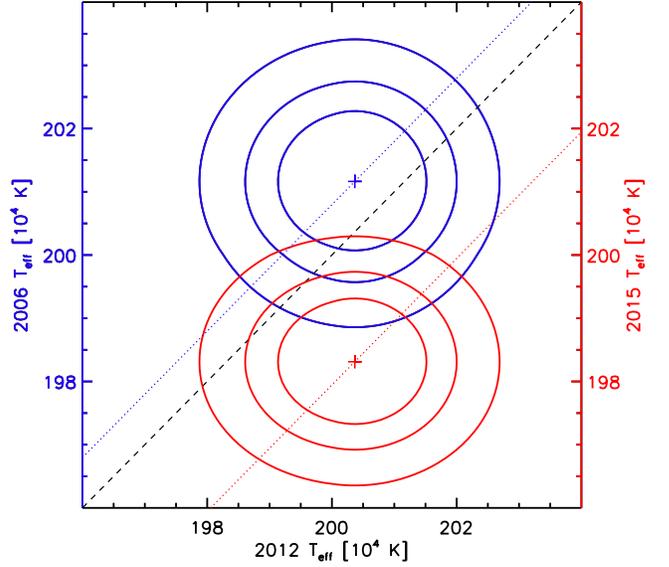}}
\caption{The temperature confidence contours (68\,\%, 90\,\%, 99\,\%) for the fit of all subarray mode data where $N_{\rm H}$ is allowed to vary between epochs (see Table~\ref{table:carbonfits} for the fit results). The blue contours correspond to the temperatures of epochs 2006 and 2012, the red contours correspond to the temperatures of epochs 2015 and 2012. The dashed line marks the line of equal temperature values, the dotted lines indicate the offsets of the best-fit values from equal temperatures. Note that the (insignificant) temperature difference for epochs 2012-2006 is also negative.
\label{t2t1t2t3}}
\end{figure}

\begin{figure}
{\includegraphics[width=85mm]{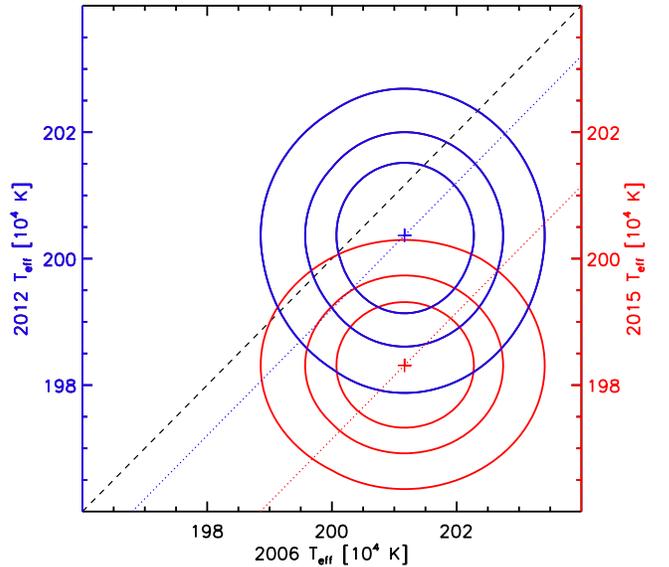}}
\caption{The temperature confidence contours (68\,\%, 90\,\%, 99\,\%) for the same fit as shown in Figure~\ref{t2t1t2t3}, but with the $x$-axis showing the values from 2006. Hence, the blue and red contours correspond to time differences of 5.5 and 8.5\,years. The dashed and dotted lines show the line of equal temperature values and offsets of the best-fit values. As demonstrated in Figure~\ref{t2t1t2t3}, the last three years contribute most to the offset of the red contours (2006-2015).
\label{t1t2t1t3}\vspace{-0.5cm}}
\end{figure}

Comparing the temperatures between all epochs, 
we find that all temperatures are statistically consistent with no change when $N_H$ is allowed to vary between the epochs, see also Figures~\ref{t2t1t2t3} and \ref{t1t2t1t3}. In the case of a tied $N_H$, the temperature change between epochs 1 and 3 is at the $3.3\sigma$ level.
A similar trend (though with lower significance) was seen by P+13 for the comparison of epochs 1 and 2. Our conclusion at the time was that the fit with tied $N_{\rm H}$ is more influenced by the contamination model uncertainties than the fit with free $N_{\rm H}$ because the latter can partly compensate for the effect of an imperfect contamination model. The substantial shift of the temperature confidence contours for the previous two epochs in Figure~\ref{t1t2w2013} supports this notion (the contours for the variable $N_{\rm H}$ are less correlated, i.e., ``rounder'', and significance levels are less affected). 
There is an interesting difference in comparison to our previous work though: Allowing $N_{\rm H}$ to vary led to a smaller difference of the best-fit temperatures in P+13, while it slightly increases the difference with the new calibration.
Since the uncertainties are larger than in the case of the tied $N_{\rm H}$ fit, the temperature differences remain, however, insignificant.\\
    
\begin{figure}
{\includegraphics[width=85mm]{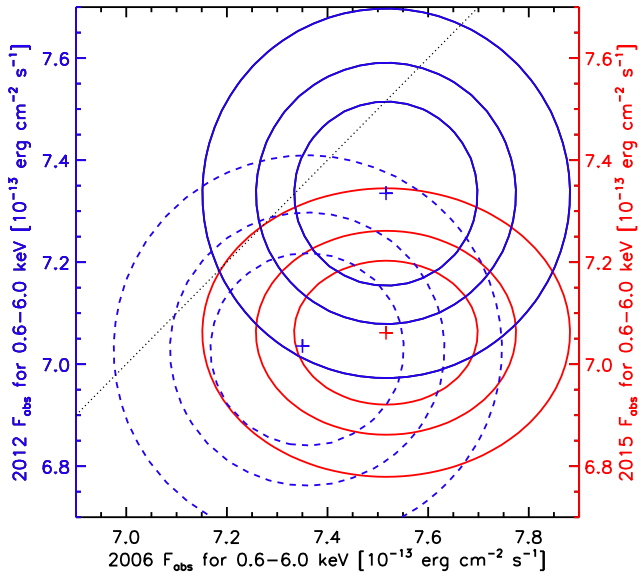}}
\caption{The confidence contours (68\,\%, 90\,\%, 99\,\%) of the absorbed fluxes in the energy range 0.6\,keV to 6.0\,keV for the fit of all subarray mode data when $N_{\rm H}$ is allowed to vary between epochs; see Table~\ref{table:carbonfits}.
The contours of the 2006-2012 data are shown in blue, those of the 2006-2015 data are shown in red. The dashed blue contours show the result obtained with CALDB 4.5.5.1 for comparison.
The straight dotted line is the line of equal fluxes. 
\label{fig:cfluxfree13}}
\end{figure}

Figure~\ref{t2t1t2t3} shows the temperature differences with respect to the second epoch, while  Figure~\ref{t1t2t1t3} shows the differences with respect to the first epoch.
The offsets of the best-fit values (dotted lines) from the line of equal temperatures (dashed) in Figure~\ref{t2t1t2t3} demonstrate that the temperature appears to `drop' slower during the first 5.5\,yr than in subsequent 3.0 years. 
Taken at the face value, such behavior would mean a rather unlikely cooling acceleration on a time scale of three years.
None of the offsets are significant, but their different values could indicate that there may still be problems with the ACIS filter calibration, e.g., for the most recent \emph{Chandra} data.
A similar interpretation can be applied to the absorbed fluxes, the values of which change barely during the first 5.5\,yr, but `drop' during the last three years (Table~\ref{table:carbonfits} and Figure~\ref{fig:cfluxfree13}).\\

\begin{figure}[h!]
{\includegraphics[width=87mm, bb=20 14 682 542, clip]{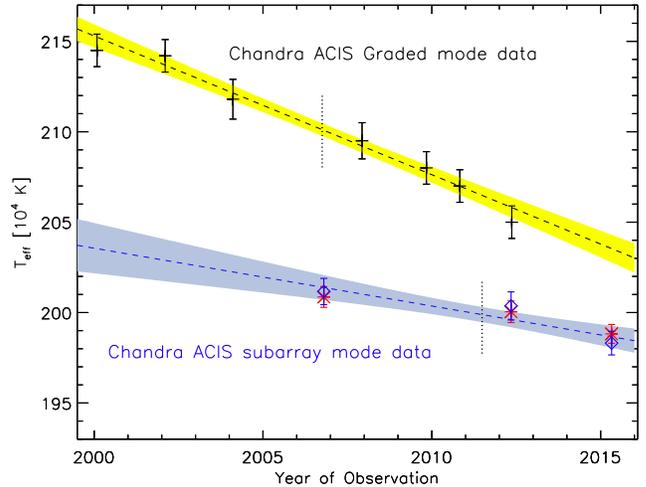}}
\caption{Temperature change over time. All errors in this plot are $1\sigma$ errors. The black crosses mark the temperatures as reported by E+13 (their Table 2), who derived them from piled-up data telemetered in Graded mode using carbon atmosphere models ($M_{\rm NS}=1.62$\,M$_{\odot}$ and $R_{\rm NS}=10.19$\,km) with the same fixed $N_{\rm H}$ for all observations.
The black dashed line and the yellow area indicate the results of a linear regression fit and its $1\sigma$ error to the E+13 data points if we choose the average of their observing epochs as reference time, $t_{\rm E13,0}=2006.75$ (dotted vertical black line), see also P+13.
Our fit results for the subarray data from Table~\ref{table:carbonfits} for a carbon atmosphere model with  similar gravitational parameters ($M_{\rm NS}=1.647$\,M$_{\odot}$ and $R_{\rm NS}=10.33$\,km) are marked with red star points (same $N_{\rm H}$ in all epochs) and blue diamond points (different $N_{\rm H}$ in 2006, 2012 and 2015). 
The blue dashed line and the blue area indicate the results of a linear regression fit and its $1\sigma$ error to the data points where $N_{\rm H}$ is allowed to vary. The chosen reference time (average time of the time span covered by the subarray observations, $t_{\rm sa, 0}=2011.49$) is marked with a dotted vertical black line.
Note that there is \emph{no significant} (i.e., $>3\sigma$) temperature decrease (the slope is $(-0.32 \pm 0.12)  \times 10^{4}$\,K\,yr$^{-1}$).
\label{tempfit}}
\end{figure}

In order to obtain an average yearly temperature change rate for the temperature data points, P+13, carried out standard least-square fits to a straight line (e.g., \citealt{Bevington2003}), $T_{\rm eff}= T_0 + \dot{T}(t-t_0)$, where $t$ is the time of observation and  $t_0$ the reference time.
Using the piled-up data points from full-frame Graded mode presented by E+13, the derived values were $\dot{T}=-7700\pm 1300 $\,K\,yr$^{-1}$ and $T_0=(210.1 \pm 0.6)\times 10^4$\,K (90\% confidence levels, $\chi^2_{\nu}=0.41$ for $\nu=5$\,dof, $t_{\rm E13,0}=2006.75$), indicated by the yellow area in Figure~\ref{tempfit}.
These values correspond to a characteristic cooling time, $\tau_{\rm cool}= T_0/(-\dot{T})=270^{+60}_{-40}$\,yr.
Our new subarray data enable us to carry out similar fits.
If $N_{\rm H}$ is allowed to vary between epochs, we derive a slope
$\dot{T}=-3200\pm 1900$\,K\,yr$^{-1}$ and an intercept $T_0=(199.9 \pm 0.7)\times 10^4$\,K (90\% confidence levels, $\chi^2_{\nu}=1.3$ for $\nu=1$\,dof, $t_{\rm sa, 0}=2011.49$), shown by the blue area in Figure~\ref{tempfit}. 
If $N_{\rm H}$ is the same for all epochs, the values are  $\dot{T}=-2200\pm 1500$\,K\,yr$^{-1}$, $T_0=(200.0 \pm 0.5)\times 10^4$\,K (90\% confidence levels, $\chi^2_{\nu}=1.6$ for $\nu=1$\,dof, $t_{\rm sa, 0}=2011.49$). 
Whether $N_{\rm H}$ is allowed to vary or not, the values of the slopes are all below their $3\sigma$ uncertainties, i.e., the temperature decrease is statistically insignificant.
The conservative $3\sigma$ upper limits on the temperature change, $-\dot{T}<6700$\,K\,yr$^{-1}$ (varying $N_{\rm H}$) and $-\dot{T}<4900$\,K\,yr$^{-1}$  (tied $N_{\rm H}$) correspond to $<3.3$\% and $<2.4$\% change in 10 years starting from the 2006 temperature, or lower limits of the characteristic cooling times of 300\,yr and 410\,yr, respectively.

\section{Conclusions}
The CCO in Cas\,A does not show statistically significant temperature or flux changes if one employs the spectra obtained in three epochs of the \emph{Chandra} ACIS subarray mode observations which are best suited to such spectral analysis. This conclusion holds for carbon atmosphere as well as hydrogen atmosphere models.
An updated calibration, in particular an updated model of the contaminant on the ACIS optical blocking filter, reduced apparent temperature or flux differences for the 2006 and 2012 data to less than $1\sigma$. This confirms the hypothesis by P+13 that an imperfect ACIS contamination model affected the previous findings. 
While the 2015 subarray data on the CCO imply lower temperature and flux values than in 2012, the overall cooling rates still remain consistent with 0.
The current conservative $3\sigma$ \emph{lower} limits on the characteristic cooling times of 300\,yr (varying $N_{\rm H}$) and 410\,yr (tied $N_{\rm H}$) exclude the 
previously reported rapid cooling by HH10 and E+13 ($\tau_{\rm cool}= 270$\,yr), 
which was inferred from data obtained using the whole ACIS chip in Graded mode.
Because those data of the Cas A CCO are impacted by a substantial pileup effect (distorting the spectra), while the subarray data show negligibe pileup, the 
theoretical results based on those data (e.g., \citealt{Negreiros2018,Burgio2018,Taranto2016,Grigorian2016,Page2011, Shternin2011}) should be revised, and the new result should be used for any theoretical constraints in neutron star cooling models.

\facility{Chandra (ACIS)}\\
\software{CIAO (v4.9; \citealt{Fruscione2006}), XSPEC (v12.8.2; \citealt{Arnaud1996})}

\acknowledgments
We thank the referee for helpful remarks.
We would also like to thank the \emph{Chandra} Director, B. Wilkes, for approving the Cas A DDT observations. We are indebted to A. Bogdan and P. Plucinsky for clarifications and help regarding the old and new models of the ACIS contamination.
We are also thankful for fruitful discussions with P. Broos, C. Heinke, W. Ho and D. Patnaude regarding the Cas A data.

The scientific results reported in this article are based on observations made by the \emph{Chandra} X-ray Observatory.
Support for this work was provided by the National Aeronautics and Space Administration through Chandra Awards G05-16068 and AR6-17008, issued by the Chandra X-ray Observatory Center, which is operated by the Smithsonian Astrophysical Observatory for and on behalf of the National Aeronautics Space Administration under contract NAS8-03060.
Support for this work was also provided by the ACIS Instrument Team contract SV4-74018 issued by the Chandra X-ray Observatory Center, which is operated by the Smithsonian Astrophysical Observatory for and on behalf of NASA under contract NAS8-03060.

This research has made use of SAOImage DS9, developed by
SAO, and SAO/NASA's Astrophysics Data System Bibliographic Services.

\bibliographystyle{aasjournal}
\bibliography{Casa}

\end{document}